\newcommand{\emet}{{\em et al.}}

\def\abstract{\subsection*{{\it  Abstract}}}

 \def\@oddfoot{\rm\rightmark \hfil \thepage \hfil}
 \def\@evenfoot{\@oddfoot}

\def\thebibliography#1{
  \section{  {\normalsize \bf REFERENCES} 
  \@mkboth {REFERENCES}{REFERENCES}}\list
 {[\arabic{enumi}]}{\settowidth\labelwidth{[#1]}
\leftmargin\labelwidth
 \advance\leftmargin\labelsep\itemsep0pt \frenchspacing
 \usecounter{enumi}}
 \def\newblock{\hskip .11em plus .33em minus -.07em}
 \sloppy
 \sfcode`\.=1000\relax}

\def\notbj#1{\mathpalette\c@ncel#1}
\def\diracs#1{\mathpalette\c@ncel#1}
\def\@startsection#1#2#3#4#5#6{\if@noskipsec \leavevmode \fi
   \par \@tempskipa #4\relax
   \@afterindenttrue
   \ifdim \@tempskipa <\z@ \@tempskipa -\@tempskipa \fi
   \if@nobreak \everypar{}\else
     \addpenalty{\@secpenalty}\addvspace{\@tempskipa}\fi \@ifstar
{\@ssect{#3}{#4}{#5}{#6}}%
{\@dblarg{\@sect{#1}{#2}{#3}{#4}{#5}{#6}}}}

\newcount\@tempcntc
\def\@citex[#1]#2{\if@filesw\immediate\write\@auxout{\string\citation{#2}}\fi
  \@tempcnta\z@\@tempcntb\m@ne\def\@citea{}\@cite{\@for\@citeb:=#2\do
    {\@ifundefined
       {b@\@citeb}{\@citeo\@tempcntb\m@ne\@citea\def\@citea{,}{\bf ?}\@warning
       {Citation `\@citeb' on page \thepage \space undefined}}%
    {\setbox\z@\hbox{\global\@tempcntc0\csname b@\@citeb\endcsname\relax}%
     \ifnum\@tempcntc=\z@ \@citeo\@tempcntb\m@ne
       \@citea\def\@citea{,}\hbox{\csname b@\@citeb\endcsname}%
     \else
      \advance\@tempcntb\@ne
      \ifnum\@tempcntb=\@tempcntc
      \else\advance\@tempcntb\m@ne\@citeo
      \@tempcnta\@tempcntc\@tempcntb\@tempcntc\fi\fi}}\@citeo}{#1}}
\def\@citeo{\ifnum\@tempcnta>\@tempcntb\else\@citea\def\@citea{,}%
  \ifnum\@tempcnta=\@tempcntb\the\@tempcnta\else
   {\advance\@tempcnta\@ne\ifnum\@tempcnta=\@tempcntb \else \def\@citea{--}\fi
    \advance\@tempcnta\m@ne\the\@tempcnta\@citea\the\@tempcntb}\fi\fi}
\def\citenum#1{{\def\@cite##1##2{[##1]}\cite{#1}}}

     \def\thefootnote{\mbox{\noindent$\fnsymbol{footnote}$}}
      \long\def\@makefntext#1{\noindent$^{\@thefnmark}$#1}

\def\maketitle{\par
 \begingroup
 \def\thefootnote{\fnsymbol{footnote}}
 \def\@makefnmark{\mbox{$^\@thefnmark$}}
 \@maketitle
 \@thanks
 \endgroup
 \setcounter{footnote}{0}
 \let\maketitle\relax
 \let\@maketitle\relax
 \gdef\@thanks{}\gdef\@author{}\gdef\@title{}\let\thanks\relax}
\def\@maketitle{\vspace*{0.2cm} 
{\hsize\textwidth
 \linewidth\hsize \centering
 {\normalsize \bf \@title \par} \vskip 0.3cm {\normalsize  \@author \par}}}

\def\thefootnote{\mbox{\noindent$\fnsymbol{footnote}$}}
\long\def\@makefntext#1{\noindent$^{\@thefnmark}$#1}

\def\thesection{\Roman{section}}
\def\thesubsection{\Alph{subsection}}

\def\section#1{
  \begin{center}\@startsection{section}{1}{\z@}
  {1.5ex plus 0.5ex minus 1.2ex}{1.3ex plus .1ex}
  {\normalsize\bf}{#1}\end{center}}

\def\subsection{\@startsection{subsection}{2}{\z@}{1.5ex plus 1ex minus
   .2ex}{1.3ex plus .1ex}{\normalsize}}

\def\subsubsection{\@startsection{subsubsection}{2}{\z@}{1.5ex plus 1ex minus
   .2ex}{1.3ex plus .1ex}{\normalsize}}

\def\@sect#1#2#3#4#5#6[#7]#8{
    \ifnum #2>\c@secnumdepth \def\@svsec{}
    \else
      \refstepcounter{#1}\edef
      \@svsec{\ifnum #2=1 \@sectname\fi \csname the#1\endcsname.\hskip 1em }
    \fi
    \@tempskipa #5\relax
    \ifdim \@tempskipa>\z@
      \begingroup #6\relax
      \@hangfrom{\hskip #3\relax\@svsec}{\interlinepenalty \@M #8\par}
      \endgroup
      \csname #1mark\endcsname{#7}\addcontentsline
      {toc}{#1}{\ifnum #2>\c@secnumdepth \else \protect\numberline{\csname
the#1\endcsname}\fi #7}
    \else
      \def\@svsechd{#6\hskip #3\@svsec #8\csname #1mark\endcsname
      {#7}\addcontentsline {toc}{#1}
      {\ifnum #2>\c@secnumdepth \else \protect\numberline{\csname
the#1\endcsname}\fi #7}}
    \fi
    \@xsect{#5}}

\def\@sectname{}

\newif\if@numbersec \@numbersectrue
\def\appendix{\par\clearpage
  \setcounter{section}{0}
  \setcounter{subsection}{0}
  \@addtoreset{equation}{section}
  \@ifstar{\def\@sectname{Appendix \ }\@numbersecfalse}
          {\def\@sectname{Appendix \ }\@numbersectrue}
  \def\theequation{\thesection\arabic{equation}}
  \def\thesection{\Alph{section}}
  \def\thesubsection{\arabic{subsection}}}

\def\thefigures#1{\par\clearpage\section*{Figures\@mkboth
  {FIGURES}{FIGURES}}\list
  {Fig. {\arabic{enumi}}.}{\labelwidth\parindent\advance\labelwidth -\labelsep
      \leftmargin\parindent\usecounter{enumi}}}

\def\thetables#1{\par\clearpage\section*{Tables\@mkboth
  {TABLES}{TABLES}}\list
  {Table {\Roman{enumi}}.}{\labelwidth-\labelsep
      \leftmargin0pt\usecounter{enumi}}}

\def\@sect#1#2#3#4#5#6[#7]#8{\ifnum #2>\c@secnumdepth
     \def\@svsec{}\else
     \refstepcounter{#1}\edef\@svsec{\ifnum #2=1 \@sectname\fi
        \csname the#1\endcsname.\hskip 1em }\fi
     \@tempskipa #5\relax
      \ifdim \@tempskipa>\z@
        \begingroup #6\relax
          \@hangfrom{\hskip #3\relax\@svsec}{\interlinepenalty \@M #8\par}
        \endgroup
       \csname #1mark\endcsname{#7}\addcontentsline
         {toc}{#1}{\ifnum #2>\c@secnumdepth \else
                      \protect\numberline{\csname the#1\endcsname}\fi
                    #7}\else
        \def\@svsechd{#6\hskip #3\@svsec #8\csname #1mark\endcsname
                      {#7}\addcontentsline
                           {toc}{#1}{\ifnum #2>\c@secnumdepth \else
                             \protect\numberline{\csname the#1\endcsname}\fi
                       #7}}\fi
     \@xsect{#5}}

\def\@sectname{}

\def\eqnarray{\stepcounter{equation}\let\@currentlabel=\theequation
\global\@eqnswtrue
\global\@eqcnt\z@\tabskip\@centering\let\\=\@eqncr
$$\arraycolsep\z@
\halign to \displaywidth\bgroup\@eqnsel\hskip\@centering
  $\displaystyle\tabskip\z@{##}$&\global\@eqcnt\@ne
  \hskip 2\arraycolsep \hfil${{}##{}}$\hfil
  &\global\@eqcnt\tw@ \hskip 2\arraycolsep $\displaystyle\tabskip\z@{##}$\hfil
   \tabskip\@centering&\llap{##}\tabskip\z@\cr}

\makeatletter
    \def\thefootnote{\mbox{\noindent$\fnsymbol{footnote}$}}
\makeatother

\begin{document}

\title{\bf Large Angle K$^+$-Deuteron Scattering\\
as a Probe for Meson Exchange Effects}

\vspace{2cm}

\author{David R. Harrington\footnote{Permanent Address: Department of Physics
and Astronomy, Rutgers University,
P.O. Box 849, Piscataway, NJ 08855-0849, USA\\
Electronic address: harrington@ruthep.rutgers.edu}\\
{\em TRIUMF, 4004 Wesbrook Mall, Vancouver, BC, Canada V6T 2A3}}

\maketitle

\vspace{1cm}

\setlength{\baselineskip}{3ex}

\begin{center}
{\bf Abstract}
\end{center}

Meson exchange contributions have been suggested as one class of medium effects
which might explain the continuing discrepancy between experimental results and
theoretical predictions for K$^+$-nucleus scattering.  These exchange effects
are negligible for near-forward K$^+$-deuteron scattering because of the large
size of the deuteron, but at large angles they might be expected to be
significant compared to the weak double scattering.  Detailed calculations,
however, show that the quadrupole and magnetic contributions from single
scattering are large enough to dominate both double scattering and exchange
contributions even at large momentum transfers.  The exchange contributions
become significant only if the target deuteron is highly aligned so that the
single scattering contributions are greatly reduced.

\newpage
\setlength{\baselineskip}{2.6ex}

\section{INTRODUCTION}

Positive kaons interact very weakly with nucleons: the total cross sections are
less than 19 mb for laboratory energies in the 600-1000 MeV range\cite{A}.  The
corresponding mean free paths in nuclear matter are greater than, or comparable
to, the sizes of even heavy nuclei, so that positive kaons, unlike other
strongly interacting probes, can penetrate the outer layers of large nuclei and
reach the dense central regions.  The multiple scattering expansion for these
interactions is also expected to converge rapidly, with single and double
scattering accounting for almost all of the total scattering amplitude\cite{B}.

It was originally expected that the K$^+$ meson would be an ideal probe to
study the nuclear interior, with the scattering theory under good control.
Accurate experiments\cite{C} have shown, however, that there are small but
definite discrepancies between calculations and observations, the most
definitive being in the ratios R\cite{B} of the total cross sections for
scattering from heavier nuclei to that for scattering from a deuteron.  It now
seems likely that this discrepancy is a medium effect\cite{B,D}, due to the
fact that the K$^+$ {\it is} sensitive to the dense interior of nuclei.
Unfortunately the exact nature of the medium effect (or effects) is still
obscure.  Several different types of medium effects have been suggested,
falling into two broad classes.  In the first the possibility that the
effective masses of the target nucleons or the  exchanged mesons might differ
from their `vacuum' values is considered\cite{E,F}. (At least some of these
effects are closely related to possible complications in the nuclear wave
functions.)  Although the original calculations gave corrections tending to
increase R, improving the agreement with experiments, a recent paper\cite{G}
has found a significant effect in the opposite direction.

The second class of medium corrections, which would also be expected to be more
important in the nuclear interior than in the less-dense outer regions, is that
due to the interaction of the K$^+$ with mesons being exchanged between
nucleons, with pi-mesons expected to be most important.  Again these
calculations are difficult since the exchanged mesons are well off shell, but
the resulting corrections to R seem to be in the right direction
\cite{H,I,J,K}.  In this paper we want to investigate whether elastic
scattering of the K$^+$ from a deuteron at fairly large angles can be used to
probe these corrections.

We were led to consider this possibility because of the following argument:
Elastic scattering from a deuteron has single and double scattering
contributions\cite{L}.  The single scattering falls off rapidly with increasing
momentum transfer because it is unlikely that the deuteron can remain together
if only one of its constituents is given a large kick. Therefore double
scattering is expected to dominate at large momentum transfer because if the
two nucleons are given similar kicks they will have a significant chance of
remaining bound.  The major contribution to double scattering is simply
successive elastic scattering from the two nucleons, but the meson exchange
contributions of the type discussed above can also be present at large angles
because they also can share the momentum transfer between the two nucleons.
These exchange contributions should be particularly important compared to the
elastic double scattering for the case of K$^+$ scattering because of the
weakness of the K$^+$-nucleon elastic interaction.  (The K$^+$-pion interaction
is not particularly weak, and is in fact dominated by the K$^*$(892) resonance
at high enough energy.)  It thus seemed that there might be a chance that large
angle K$^+$-deuteron elastic scattering would be particularly sensitive to
meson exchange contributions.

Complications arise, however, because of the D-wave component of the deuteron's
wave function and the spin dependence of the K$^+$-N scattering
amplitudes\cite{L}.  These introduce into the scattering amplitudes deuteron
spin dependent terms  proportional to the deuteron's quadrupole and magnetic
form factors.  These  form factors are small for near-forward scattering but
become larger than the spherical form factor at large momentum transfers where
the double scattering begins to be important.  As we shall see below, the
contribution of the quadrupole and magnetic form factors to single scattering
dominates both the double scattering and meson exchange contributions, greatly
reducing the sensitivity of the cross-section to the exchange contributions,
even at large angles. (For other strongly interacting projectiles the double
scattering is large enough to dominate the quadrupole and magnetic terms at
large angles.  In $\pi$d scattering, for example, the quadrupole term is
important mainly in the single-double transition region, while the magnetic
term is almost negligible\cite{L}.)

There is a way to avoid this reduction in sensitivity, however.  If the target
deuteron is aligned, then the quadrupole contribution can interfere with the
spherical contribution instead of contributing incoherently as for the case of
scattering from unpolarized deuterons.  For certain choices of alignments the
single and elastic double scattering nearly cancel for a range of momentum
transfers and the differential cross section would have a fairly deep minimum
were it not for the exchange and magnetic contributions.  Even the latter can
be eliminated if the deuteron is properly aligned along the normal to the
scattering plane.  For large momentum transfers the differential cross sections
for scattering from the aligned deuterons then becomes quite sensitive to the
exchange contribution.  It thus seems possible to test various prescriptions
for the exchange amplitude, but only under quite challenging experimental
conditions.

We begin, in Sec.~II, by reviewing the expressions for the single scattering,
double scattering, and pion exchange contributions to the K$^+$- deuteron
scattering amplitude, including the contributions from the spin- dependence of
the K$^+$-nucleon amplitudes, and the D-wave component of the deuteron wave
function.  The pion exchange contribution depends critically on the off-shell
pion-K$^+$ scattering amplitude, which we discuss in more detail in Sec.~III.
Numerical results are presented in Sec.~IV, where it is shown that the
sensitivity of the differential cross section to the exchange contribution can
be greatly enhanced if the target deuteron is in a properly chosen alignment
state.  We conclude, in Sec.~V, with a brief discussion of the experimental
difficulties implied by our results.

\section{SCATTERING AMPLITUDE}

We shall use Glauber theory to estimate the single and elastic double
scattering contributions to the K$^+$-deuteron scattering amplitudes.  There
may be some problems with the accuracy of this approximation due the short
range of the K$^+$-nucleon interaction, but it produces simple formulas which
should at least give reasonable estimates of the various contributions in the
region of momentum transfer of interest here.  The single scattering amplitude,
corresponding to Fig.~1(a),  at momentum transfer $\Delta$, is then
\begin{eqnarray}                       
F^{(1)} (\Delta) = 2\overline{f}(\Delta) [S_0 (\Delta/2) -
[3(J\cdot\hat{\Delta})^2-2] S_2 (\Delta/2)] + 2i\overline{g} (\Delta) S_M
(\Delta/2) (J\cdot\hat{n})\,\, ,
\end{eqnarray}
where

\newpage
\begin{eqnarray}
\overline{f} (\Delta) &=& (f_0 (\Delta) +3 f_1 (\Delta))/4
\nonumber \\
\overline{g} (\Delta) &=& (g_0 (\Delta) +3 g_1 (\Delta))/4\,\, .
\end{eqnarray}
Here $f_I$  and $g_I$ are the kaon-nucleon scattering spin-independent and
spin-dependent amplitude for isospin I=0 or 1, respectively, while $S_0$,
$S_2$ and  $S_M$  are the deuteron's spherical, quadrupole and magnetic form
factors, and J is the deuteron's spin.  The double scattering amplitude of
Fig.~1(b) is given by the integral
\begin{eqnarray}                  
F^{(2)} (\Delta) = (i/2\pi) \int d^2\delta S_0 (\Delta)
G(\Delta,\delta)\,\, ,
\end{eqnarray}
where
\begin{eqnarray}                   
G(\Delta,\delta) &=& [3 f_1 (\delta_+)  f_1 (\delta_-) +3 f_1
(\delta_+) f_0 (\delta_-)\nonumber \\
\mbox{}& &+3 f_0 (\delta_+)  f_1 (\delta_-) -f_0
(\delta_+)  f_0 (\delta_-)]/8\,\, .
\end{eqnarray}
The contribution to the elastic scattering due to the interaction of the kaon
with a pion exchanged between the two nucleons, as shown in Fig.~1(c), is given
by\cite{M}
\begin{eqnarray}                    
F^{(3)} (\Delta) = (3g^2/32\pi m^2k_{lab}) \int
(d^3\delta/(2\pi)^3)\frac{S_0(\delta)
\sigma_1\cdot \delta_+\sigma_2\cdot\delta_- M_{K\pi}}
{(\delta^2_++\mu^2)(\delta^2_-+\mu^2)}\,\, .
\end{eqnarray}

Here, and in Eq.~(4), $\delta_{\pm}=\delta\pm \Delta/$2,
while M$_{K\pi}$ is
the isospin zero invariant amplitude for K-$\pi$ scattering.  In the
expressions
for F$^{(2)}$ and F$^{(3)}$ we have ignored the D-wave component of the
deuteron
wave function, simplifying the expressions without much loss in accuracy.

Because of the spin operators J in these expression the differential cross
section will be dependent on the spin state of the target deuteron.
(Alternatively, if the target deuterons are unpolarized the recoil deuterons
will be polarized.)

The largest uncertainties in our results are associated with the exchange
contributions of Eq.~(5), which we shall discuss in more detail in the next
section.

\section{EXCHANGE CONTRIBUTION}

The invariant amplitude M$_{K\pi}$ in Eq.~(5) is required for off-shell pions.
In fact, in the static approximation used to derive Eq.~(5), the pion mass
variables take the values $-(\delta_\pm)^2$, which are never positive.  For
small values of the Mandelstam variables
\begin{eqnarray}                         
s &=& m^2_k + \Delta^2/4 + \delta^2 + 2k\cdot \delta\nonumber \\
u &=& m^2_k + \Delta^2/4 + \delta^2 - 2k\cdot \delta\nonumber \\
t &=& -\Delta^2\,\, ,
\end{eqnarray}
estimates of M$_{K\pi}$ can be made using soft-pion techniques\cite{N,O}, but
it is unlikely that currently available results will be adequate over the
variable region required to evaluate the integral in\cite{E}.  For very
energetic K$^+$'s, M$_{K\pi}$ will be dominated by the K$^\ast$(892) resonance,
and the meson exchange contribution will evolve into a piece of the inelastic
double scattering amplitude.  Since the purpose of this work is to see how
sensitive large-angle K$^+$-deuteron scattering is to exchange contributions,
we shall simply use the prescription of Ref.~10, which includes both soft-pion
and resonance contributions.  We shall not, however, include the $\pi$KKNN
contact interaction\cite{P} mentioned in Ref.~10.  This interaction, required
to cancel the one-pion exchange contribution to KN$\to$KN$\pi$ at the Adler
point, would tend to weaken the exchange contribution.

Because of the spin operators in Eq.~(5), F$^{(3)}$ has terms proportional to
$(J\cdot \Delta)^2$ and $(J\cdot k)^2$, the latter due
only to the k-dependence of M$_{K\pi}$, which comes only from the resonance
terms.  A $(J\cdot \Delta)^2$
term also occurs in the quadrupole  contribution to single scattering, but the
$(J\cdot k)^2$ operator appears only in the exchange contribution.

\section{NUMERICAL RESULTS}

As mentioned above, the exchange contribution to K$^+$-deuteron scattering
has no chance of being significant except at large momentum transfers where
the single scattering has decreased and may at least partially cancel with
double scattering.  In order to reach these large momentum transfers without
going to near-backward angles, where the Glauber approximation would clearly
be invalid, we must go to fairly high energies: all of the calculations below
are done at a laboratory momentum of 970~MeV/c.  (The only available
experimental data\cite{Q} at energies close to this are confined to small
angles.)

In evaluating the formulas in Sec.~II we have used the K$^+$N amplitudes
of\linebreak[4]
Hashimoto\cite{A} and the fits of Ref.~18 to the deuteron form factors for the
Paris potentials.  The  integral in\cite{C} is weakly dependent on the K$^+$N
amplitudes outside of the physical region.  The results below assume a constant
extrapolation, but are not much different than if the amplitudes were assumed
zero outside the physical region.  The amplitudes F$^{(2)}$ and F$^{(3)}$
depend
smoothly on the momentum transfer $\Delta$.  The integrals in Eqs.~(3) and (5)
were evaluated for a range of $\Delta$'s and the results fitted to polynomials
which were then used in the numerical evaluations below.

Although the kinematics of K$^+$-deuteron scattering is identical to that of
$\pi$-deuteron scattering except for the meson masses, the weakness of the
K$^+$-nucleon interaction compared to the $\pi$-nucleon interaction means that
some features of the cross sections for the two interactions are quite
different. Figure~1 shows the relative sizes of the various contributions to
the differential cross section as a function of -t, the square of the momentum
transfer.  (The various terms can in some cases interfere with each other when
they are combined, so the total differential cross section is not the sum of
the terms shown.)   The double scattering amplitude in K$^+$d scattering is so
small, for example, that it is dominated by the quadrupole contribution to
single scattering at large angles, in contrast to $\pi$d scattering\cite{L}
where the quadrupole contribution is important only in the transition region,
where the spherical single scattering and double scattering amplitudes nearly
cancel.  Even the magnetic contribution to single scattering, arising from the
spin- dependent piece of the of the K$^+$N amplitude, is greater than the
double scattering amplitude, in contrast to the $\pi$d case where it is almost
negligible\cite{L}.  The exchange contribution is smaller than the double
scattering contribution by a factor of roughly five at large momentum
transfers, but of course both are masked by the single scattering terms.  This
is seen more clearly in Fig.~2 where the differential cross section for
scattering from unpolarized deuterons is shown with and without the
contribution from the exchange contribution.  Even at the largest momentum
transfers shown the exchange contribution adds only about 40\% to the
differential cross section.

It is well known, however, that the shape of the differential cross section for
scattering from deuterons depends strongly on the polarization of the target
deuteron\cite{S}, and it is now possible to produce highly aligned deuteron
targets\cite{T}.  If the deuteron has alignment 1 (an equal mixture of
m=$\pm$1,with no m=0 component) along the momentum transfer direction, then the
spherical and quadrupole contributions to single scattering become coherent,
and tend to cancel near -t=8.5 fm$^{-2}$.  This increases the relative
importance of the exchange contribution somewhat, as shown in Fig.~3, but its
effect is still diluted considerably by the presence of the magnetic single
scattering term.  This term can, however, be eliminated completely if the
target deuteron has alignment -2 (a pure m=0 state) along the normal to the
scattering plane.  In this case, as shown in Fig.~4,  the differential cross
section above -t=8 fm$^{-2}$ is changed decisively by the exchange
contribution:  the deep dip is partially filled in, and beyond the dip region
the cross section is increased by a factor of about 2 because of the
considerable cancellation between the double and single scattering terms.  The
sensitivity to the exchange contribution decreases rapidly, unfortunately, as
the alignment increases from this extreme value, as can be seen by comparing
Fig.~4 with Fig.~5, which shows the differential cross section for an alignment
of -1.9 along the same direction.

\section{DISCUSSION}

The results above show that it is in principle possible to use large angle
K$^+$d scattering experiments to probe the pion exchange contribution and to
test whether our estimates for the off-shell K$^+\pi$ amplitude are reasonable.
Unfortunately the experiments required will be quite difficult.  The exchange
contributions are completely negligible until the momentum transfer squared is
larger than 8 fm$^{-2}$ or so, where the differential cross section has
decreased by more than two orders of magnitude from its forward value.
Furthermore, in order to maximize the sensitivity to the exchange contribution,
the target deuteron must be nearly perfectly aligned, and the angular
resolution must be quite high since the dip in the differential cross section,
where the sensitivity is greatest, is quite narrow.

It should be emphasized that the calculations here are of an exploratory
nature.  The exchange contribution in particular is representative only, and we
have made no attempt to improve the prescription of Ref.~10.  In this energy
range the contribution of the resonant piece of the K$^+-\pi$ amplitude is
rather small so that it might well be possible to use extended soft-meson or
chiral perturbation theory to improve the calculation.  It appears, however,
that because of the presence of the KNNK$\pi$ contact term, a full calculation
will have to involve the nucleons as well as the mesons, and thus be quite
difficult.

\begin{center}
{\bf ACKNOWLEDGEMENTS}
\end{center}

The author would like to thank Harold Fearing and his colleagues in the Theory
Group at TRIUMF for their kind hospitality.  He would also like to thank Avram
Gal and Judah Eisenstein for discussions of the general K$^+$-nucleus
scattering
problem which led to this work.

\newpage

\begin{center}
{\bf Figure Captions}
\end{center}

\begin{enumerate}
\item  Diagrams (a), (b) and (c) correspond to the contributions to K$^+$-
deuteron scattering from single scattering, double scattering, and meson
exchange, respectively.  The double lines represent the deuteron, the solid
curves the nucleons, the dashed lines the kaon, and the dotted line a pion.
The circles with horizontal stripes indicate the K$^+$-nucleon amplitude, the
circle with vertical stripes the K$^+$-pion amplitude.
\item  The calculated partial contributions to the differential cross
section for K$^+$-deuteron elastic scattering at 970~MeV/c laboratory momentum
from various processes.  Some of these partial contributions can add
coherently in the total differential cross section, which is therefore not
just their sum.
\item  The calculated differential cross section for K$^+$-deuteron elastic
scattering from unpolarized deuteron targets at 970 MeV/c with (solid curve)
and without (dashed curve) the exchange contributions.  The plus signs
indicate experimental data at 890~MeV/c from Ref.~17.
\item  The same as Fig.~3 except that the target deuterons have an
alignment of 1.0 (no m=0 state) along the direction of the momentum transfer.
\item  The same as Fig.~4 except that the alignment is -2.0 (a pure m=0
state) along the normal to the scattering plane.
\item  The same as Fig.~5 except that the alignment has been changed to
-1.9, so that the deuteron spin state has a small admixture of m=$\pm$1 states.

\end{enumerate}

\end{document}